\documentclass[11pt]{article}

\usepackage[margin=1in]{geometry}
\usepackage{graphicx}
\usepackage[round]{natbib}
\usepackage{hyperref}
\usepackage{authblk}

\hypersetup{
  colorlinks=true,
  linkcolor=blue,
  citecolor=blue,
  urlcolor=blue
}

\title{kindling: A Higher-Level torch Interface for Generating, Training, and Tuning Neural Networks in R}

\author[1]{Antoine Soetewey (ORCID: 0000-0001-8159-0804)}
\author[2]{Joshua Marie}
\affil[1]{HEC Li\`ege, Universit\'e de Li\`ege, Rue Louvrex 14, 4000 Li\`ege, Belgium}
\affil[2]{Independent Researcher}

\date{03 July 2026}

\begin{document}

\maketitle

\section{Summary}

\texttt{\{kindling\}} is an R \citep{rcoreteam2025} package that provides a higher-level interface to the \texttt{\{torch\}} package \citep{falbelluraschi2025torch}, R's native implementation of PyTorch, for defining, training, and tuning neural networks. This package supports MLPs with the same topology, including the standard deep feedforward neural networks and recurrent variants (RNN, LSTM, GRU), while reducing the boilerplate typically required to write \texttt{\{torch\}} neural network architecture expression and training loops by hand.

\begin{figure}[htbp]
\centering
\includegraphics[width=0.75\textwidth]{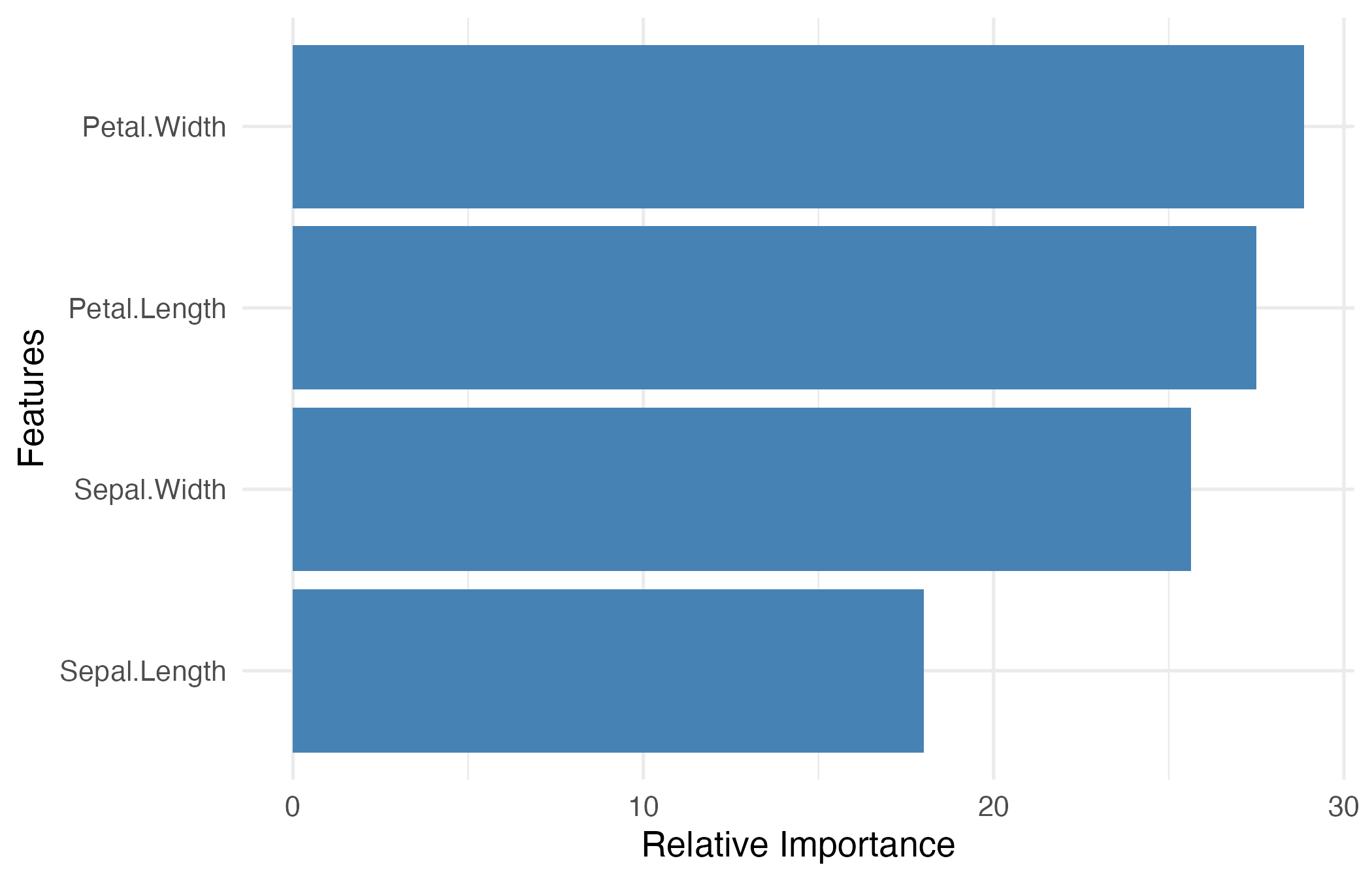}
\caption{Garson's algorithm variable-importance scores \citep{garson1991}, computed with \texttt{\{kindling\}}'s \texttt{garson()} wrapper around \texttt{\{NeuralNetTools\}} \citep{beck2018neuralnettools} for a feedforward network trained on the \texttt{iris} dataset (the same model used in the package's README ``Direct Training Interface'' example). Generated with: \texttt{model = ffnn(Species \textasciitilde{} ., data = iris, hidden\_neurons = c(10, 15, 7), activations = act\_funs(relu, softshrink[lambd = 0.5], elu), loss = "cross\_entropy", epochs = 100); garson(model, bar\_plot = TRUE)}.}
\label{fig:garson}
\end{figure}

The package is organized around three levels of abstraction. At the lowest level, \texttt{\_generator} functions (for example \texttt{ffnn\_generator()}) return unevaluated \texttt{torch::nn\_module()} expressions that users can inspect or modify directly, since \texttt{\{kindling\}} builds its models through code generation rather than opaque wrapper objects. At an intermediate level, functions such as \texttt{ffnn()} and \texttt{rnn()} train a model directly from a formula and data frame, handling data preparation, the optimization loop, and, optionally, early stopping and a validation split. At the highest level, \texttt{mlp\_kindling()} and \texttt{rnn\_kindling()} register these models as \texttt{\{parsnip\}} model specifications \citep{kuhnvaughan2026parsnip}, so they can be fit, tuned, and evaluated using the rest of the \texttt{\{tidymodels\}} ecosystem \citep{kuhnwickham2020tidymodels}: \texttt{\{recipes\}} for preprocessing, \texttt{\{workflows\}} for bundling preprocessing and modeling steps, and \texttt{\{tune\}}/\texttt{\{dials\}} for hyperparameter search over layer widths, network depth, activation functions, the output activation, the optimizer, and other architectural choices. Fitted models can also be inspected with variable-importance methods from \texttt{\{NeuralNetTools\}} \citep{beck2018neuralnettools}, implementing the algorithms of Garson \citep{garson1991} and Olden and Jackson \citep{oldenjackson2002}, and with the \texttt{\{vip\}} package. \autoref{fig:garson} shows an example of this last capability applied to the feedforward network trained in the package's own README usage example.

\section{Statement of need}

\texttt{\{torch\}} gives R users direct access to tensors, automatic differentiation, and GPU acceleration, but writing a \texttt{torch::nn\_module()}, a training loop, and the surrounding data-handling code by hand is repetitive and error-prone, particularly for users who want to compare several architectures or run a hyperparameter search rather than fit one fixed model. At the same time, R's dominant applied machine learning framework, \texttt{\{tidymodels\}}, historically has had comparatively narrow neural network support, and most existing higher-level \texttt{\{torch\}} wrappers do not expose the generated model code or integrate with the tuning and resampling infrastructure that \texttt{\{tidymodels\}} users already rely on for other model types.

\texttt{\{kindling\}} addresses this gap for R users who want deep learning to sit inside a \texttt{\{tidymodels\}} workflow rather than beside it. Because model specifications built with \texttt{mlp\_kindling()} or \texttt{rnn\_kindling()} behave like any other \texttt{\{parsnip\}} model, an analyst who already tunes a random forest or a boosted tree with \texttt{tune::tune\_grid()} and \texttt{\{rsample\}} resamples can point the same workflow at a neural network with only the model specification changed. The code generation layer additionally lets users audit or extend the underlying \texttt{\{torch\}} module instead of treating it as a black box, which is useful both for teaching and for architectures that need small custom modifications.

\section{State of the field}

Several R packages sit above \texttt{\{torch\}} to reduce this boilerplate, and each targets a different point on the tradeoff between flexibility and convenience. \texttt{\{brulee\}} \citep{kuhnfalbel2025brulee} is the official \texttt{\{tidymodels\}} package for \texttt{\{torch\}}-based models; it offers production-oriented, batteries-included implementations of linear, logistic, and multinomial regression and a multi-layer perceptron whose depth and per-layer activations (from a fixed built-in list) can be set manually but are not exposed as \texttt{\{dials\}}-tunable search dimensions the way \texttt{\{kindling\}}'s \texttt{n\_hlayers()}/\texttt{grid\_depth()} and \texttt{activations()} are; it also does not support recurrent or convolutional architectures, custom activation functions, or inspectable generated code. \texttt{\{cito\}} \citep{amesoder2024cito} emphasizes statistical inference and explainability for fully-connected networks and convolutional networks through a formula interface, with an extensive set of interpretation tools (partial dependence, accumulated local effects, bootstrap confidence intervals), but it is a standalone package rather than a \texttt{\{tidymodels\}} engine. \texttt{\{luz\}} provides a general, high-level training-loop abstraction for arbitrary \texttt{torch::nn\_module()} objects \citep{falbel2025luz}; it is architecture-agnostic and reduces boilerplate at the loop level, but does not offer a formula interface, code generation, or \texttt{\{tidymodels\}} integration.

\texttt{\{kindling\}} is positioned alongside, not in place of, these packages: it combines inspectable code generation, both feedforward and recurrent architecture families, and full \texttt{\{tidymodels\}} integration (\texttt{\{parsnip\}} specifications, \texttt{\{tune\}}/\texttt{\{dials\}} search spaces, and \texttt{\{recipes\}}/\texttt{\{workflows\}} pipelines) in one package. Its code generation layer is also more versatile than a feedforward/recurrent split suggests: the generalized \texttt{nn\_module\_generator()}/\texttt{train\_nn()} interface (currently experimental) can generate and train arbitrary sequential \texttt{\{torch\}} architectures, for example one-dimensional convolutional networks, beyond the fixed model families exposed by \texttt{\{brulee\}} and \texttt{\{cito\}}. \texttt{\{mlr3\}} \citep{lang2019mlr3} is the other major R modeling framework and is a planned integration target for \texttt{\{kindling\}}, alongside its current \texttt{\{tidymodels\}} support. In practice, the choice between these packages depends on the task: \texttt{\{cito\}} for statistical inference and explainability, \texttt{\{luz\}} for custom training loops, and \texttt{\{brulee\}} or \texttt{\{kindling\}} for production-oriented \texttt{\{tidymodels\}} workflows, where \texttt{\{kindling\}}'s own full \texttt{\{tidymodels\}} integration makes it just as production-ready as \texttt{\{brulee\}} while additionally offering architectural flexibility, broader architecture support, and inspectable generated code.

\section{Software design}

\texttt{\{kindling\}}'s central design decision is to separate code generation from execution. The \texttt{\_generator} functions build quoted \texttt{torch::nn\_module()} expressions through R's metaprogramming facilities rather than instantiating an opaque model object internally; the same expression can be printed, copied, hand-edited, and run outside \texttt{\{kindling\}}. This costs an extra layer of implementation complexity (an expression-building step that a direct wrapper would not need), not a runtime cost, since the generated code executes as ordinary \texttt{\{torch\}} once built; in exchange, a user, reviewer, or student can verify exactly what network is being fit rather than trust a black box, and can extend a generated architecture for a case the higher-level wrappers do not cover.

The three-level API (code generation, direct training, \texttt{\{tidymodels\}} engine) follows from that same trade-off, applied at increasing levels of convenience and constraint. The generalized \texttt{train\_nn()} interface dispatches via S3 methods on the class of its input (\texttt{matrix}, \texttt{data.frame}, \texttt{formula}, or \texttt{\{torch\}} \texttt{dataset}); \texttt{ffnn()} and \texttt{rnn()} are separate, architecture-specific entry points that accept a formula or \texttt{x}/\texttt{y} arguments directly rather than going through that dispatch (a possible target for consolidation once the package has matured further post-review). All three delegate preprocessing to \texttt{hardhat::mold()}/\texttt{forge()} rather than reimplementing it, so \texttt{\{kindling\}} models inherit the same predictor-role, factor-encoding, and missing-data handling as other \texttt{\{parsnip\}}/\texttt{\{recipes\}} engines. Tunable parameters (\texttt{hidden\_neurons()}, \texttt{activations()}, \texttt{output\_activation()}, \texttt{grid\_depth()}) are implemented as \texttt{\{dials\}} parameter objects dispatched over \texttt{default}/\texttt{list}/\texttt{model\_spec}/\texttt{param} classes, so they plug directly into \texttt{tune::tune\_grid()}'s search-space machinery instead of a bespoke tuning loop. This constrains \texttt{\{kindling\}} to \texttt{\{tidymodels\}} conventions, but it is precisely what lets a \texttt{\{tidymodels\}} user reuse existing resampling and tuning code for a neural network instead of learning a separate API.

Within the generated code itself, activation functions are specified through \texttt{act\_funs()}, a small embedded domain specific language (eDSL) function (akin to \texttt{\{dplyr\}}'s \texttt{pick()} function) based on R's non-standard evaluation (NSE), built with \texttt{\{rlang\}}, that accepts bracket syntax such as \texttt{softshrink[lambd = 0.5]} for parametric activations, and \texttt{new\_act\_fn()} as an escape hatch for activations with no \texttt{torch::nnf\_*()} equivalent. This trades some NSE complexity in the implementation for a call-like syntax at the user level, avoiding nested lists of function names and parameter values.

\section{Research impact statement}

\texttt{\{kindling\}} is a young package (development began with its first commit around early November 2025) and does not yet have external citations, publications, or third-party integrations to report; the evidence below documents community-readiness rather than realized adoption, as of this writing (July 2026). It is available on CRAN (initial release 0.1.0 on 2026-01-31; current release 0.3.1), with 1,434 downloads recorded since that release and 278 downloads in the last month, and the GitHub repository has 27 stars and 5 forks. Correctness is backed by a test suite of 206 \texttt{\{testthat\}} test blocks covering the code generators, the direct-training wrappers, the \texttt{\{tidymodels\}} engines, and the tuning/\texttt{\{dials\}} integration, run continuously via GitHub Actions \texttt{R-CMD-check} and tracked with Codecov.

\section{AI usage disclosure}

The authors used Claude (Anthropic) to obtain suggestions for English phrasing and to improve clarity and readability during the writing process. All content was critically reviewed and finalized by the authors.

\section{Acknowledgements}

We thank the maintainers of \texttt{\{torch\}} for R and of the \texttt{\{tidymodels\}} ecosystem, on which \texttt{\{kindling\}} is built.

\bibliographystyle{apalike}
\bibliography{paper}

\end{document}